\documentclass[
 reprint,
superscriptaddress,
 amsmath,amssymb,
 aps,
]{revtex4-2}

\usepackage[table,dvipsnames]{xcolor}
\usepackage{dcolumn}
\usepackage{bm}
\usepackage[T1]{fontenc}
\usepackage{braket}
\usepackage{amsmath}
\usepackage{amssymb}
\usepackage{nccmath}
\usepackage{mathrsfs}
\usepackage{multirow}

\usepackage{pgf}   
\usepackage{graphicx}
\let\oldsim\sim 
\renewcommand{\sim}{{\oldsim}}

\graphicspath{ {./images/} }

\newcommand{\sm}{s_\mathrm{mol}}
\newcommand{\sa}{s_\mathrm{atom}}
\newcommand{\half}{{}^1/_{2}}

\begin{document}
\title{Spectrum of Feshbach resonances in NaLi $+$ Na collisions}

\author{Juliana J. Park}
\email[]{jjpark@mit.edu}
\affiliation{Research Laboratory of Electronics, MIT-Harvard Center for Ultracold Atoms,
Department of Physics, Massachusetts Institute of Technology, Cambridge, Massachusetts 02139, USA}

\author{Hyungmok Son}
\affiliation{Research Laboratory of Electronics, MIT-Harvard Center for Ultracold Atoms,
Department of Physics, Massachusetts Institute of Technology, Cambridge, Massachusetts 02139, USA}
\affiliation{Department of Physics, Harvard University, Cambridge, Massachusetts 02138, USA}

\author{Yu-Kun Lu}
\affiliation{Research Laboratory of Electronics, MIT-Harvard Center for Ultracold Atoms,
Department of Physics, Massachusetts Institute of Technology, Cambridge, Massachusetts 02139, USA}

\author{Tijs Karman}
\affiliation{Institute for Molecules and Materials, Radboud University, Heijendaalseweg 135, 6525 AJ Nijmegen, the Netherlands}

\author{Marcin Gronowski}
\affiliation{Faculty of Physics, University of Warsaw, Pasteura 5, 02-093 Warsaw, Poland}

\author{Micha{\l} Tomza}
\affiliation{Faculty of Physics, University of Warsaw, Pasteura 5, 02-093 Warsaw, Poland}

\author{Alan O. Jamison}
\affiliation{Institute for Quantum Computing and Department of Physics \& Astronomy,
University of Waterloo, Waterloo, Ontario N2L 3G1, Canada}

\author{Wolfgang Ketterle}
\affiliation{Research Laboratory of Electronics, MIT-Harvard Center for Ultracold Atoms,
Department of Physics, Massachusetts Institute of Technology, Cambridge, Massachusetts 02139, USA}



\begin{abstract}
	
Collisional resonances of molecules can offer a deeper understanding of interaction potentials and collision complexes, and allow control of chemical reactions. Here, we experimentally map out the spectrum of Feshbach resonances in collisions between ultracold triplet ro-vibrational ground-state NaLi molecules and Na atoms over a range of 1400 G. Preparation of the spin-stretched state puts the system initially into the non-reactive quartet potential. A total of 25 resonances are observed, in agreement with quantum-chemistry calculations using a coupled-channels approach. Although the theory cannot predict the positions of resonances, it can account for several experimental findings and  provide unprecedented insight into the nature and couplings of ultracold, strongly interacting complexes. Previous work has addressed only weakly bound complexes. We show that the main coupling mechanism results from spin-rotation and spin-spin couplings in combination with the anisotropic atom-molecule interaction, and that the collisional complexes which support the resonances have a size of 30-40 $a_0$.  This study illustrates the potential of a combined experimental and theoretical approach.
\end{abstract}

\maketitle	
	
\section{Introduction}

Collisional resonances that are electromagnetically tunable have become an established tool for modifying interactions between ultracold atoms and are the key for many applications, from magnetic association of loosely bound molecules to quantum simulations \cite{chin2010feshbach, bloch2012quantum}. For ultracold molecular systems, tunable collisional resonances can control chemical reactions \cite{zhao2022quantum} and also provide microscopic information about interaction potentials and collision complexes.

In the case of cold collisions between alkali-metal \textit{atoms}, the number of resonant states remains typically small, and the resonances are usually tractable. However, in the case of cold collisions involving \textit{molecules}, due to strong and anisotropic interactions, ro-vibrational excited states can also contribute to resonant states, and therefore resonances themselves may not be well separated and are difficult to identify.
Due to the large density of states of molecular systems \cite{mayle2012statistical, christianen2019quasiclassical} it has been a challenge to perform rigorous scattering calculations and different methods of approximations have become an active field \cite{morita2019restricted, koyu2022total}. 
Despite these efforts, due to the extreme sensitivity of the low-temperature observables to details of potential energy surfaces \cite{wallis2011prospects, morita2019universal}, there has been very little success in predicting Feshbach resonances in molecular collisions. 

Rather than performing exact quantum calculations,  one feasible alternative approach is the use of simple statistical short-range models while treating the physics of long-range scattering within multichannel quantum defect theory.  This has been pursued for the description of molecular resonances \cite{mayle2012statistical, mayle2013scattering, christianen2019quasiclassical, christianen2021lossy}. However, the validity of statistical short-range models is controversial and, therefore, detailed experimental and further theoretical studies are needed.


Previously, only resonances supported by long-range states of collision complexes could be assigned to specific quantum states since the quantum numbers of the separated atoms and molecules are approximately preserved. Resonant states in collisions involving Feshbach molecules \cite{chin2005observation, knoop2009observation, zenesini2014resonant, wang2019observation} and in collisions between $^{40}\rm{K}$ and ground state $^{23}\rm{Na}^{40}\rm{K}$($X^1\Sigma^+$) \cite{yang2019observation, wang2021magnetic} were successfully analyzed in this way. Here we are addressing the challenge of more strongly interacting complexes by using, for the first time, a combined experimental and theoretical quantum chemistry approach towards this goal.  The agreement between experiment and theory validates approximations in the calculations, and the theoretical results can then be used to assign quantum numbers to resonances and to identify the microscopic mechanism of the resonant couplings. 
To keep the problem tractable, we focus on collisions between triplet ro-vibrational ground-state NaLi and Na, both prepared in the maximally spin-stretched state. Collisions of $^{23}\rm{Na}^{6}\rm{Li}$($a^3\Sigma^+$) with $\rm{Na}$ are generally chemically reactive ($^{23}\rm{Na}^{6}\rm{Li}$($a^3\Sigma^+$) $+$ $\rm{Na} \rightarrow \rm{Na}_2$$(X^1\Sigma_g^+) +\rm{Li} + \rm{heat}$). However, the chemical reaction in the fully spin-polarized atom–molecule system is strongly suppressed due to the approximate conservation of the total spin \cite{son2022control,tomza2013chemical}, and we can regard NaLi $+$ Na in the quartet potential as approximately chemically stable. The system is suitable for modeling molecular scattering resonances because of the relatively small density of states and number of electrons, and therefore more accurate quantum calculations are feasible compared to other heavier molecular systems.  Although the observed resonances involve strongly bound complexes, we can make complete assignments of spin states and total mechanical angular momentum and identify the relevant coupling terms.

In previous work, we have analyzed the lineshape of one very strong resonance in collisions between  $^{23}\rm{Na}$ and $^{23}\rm{Na}^{6}\rm{Li}$($a^3\Sigma^+$) \cite{son2022control}.  Here we report a study of collisions in both spin-stretched states over a magnetic field range of over 1400 G and report more than 20 new Feshbach resonances.  This now allows us to draw conclusions about typical features of collisional complexes. The direct comparison with quantum chemistry calculations provides major new insight into the interaction potential and coupling mechanism of the strongly interacting collision complex.



\section{Experimental protocol and results}

\begin{figure*}
 \begin{minipage}{\linewidth}
    \centering
    \includegraphics[width = 175mm, keepaspectratio]{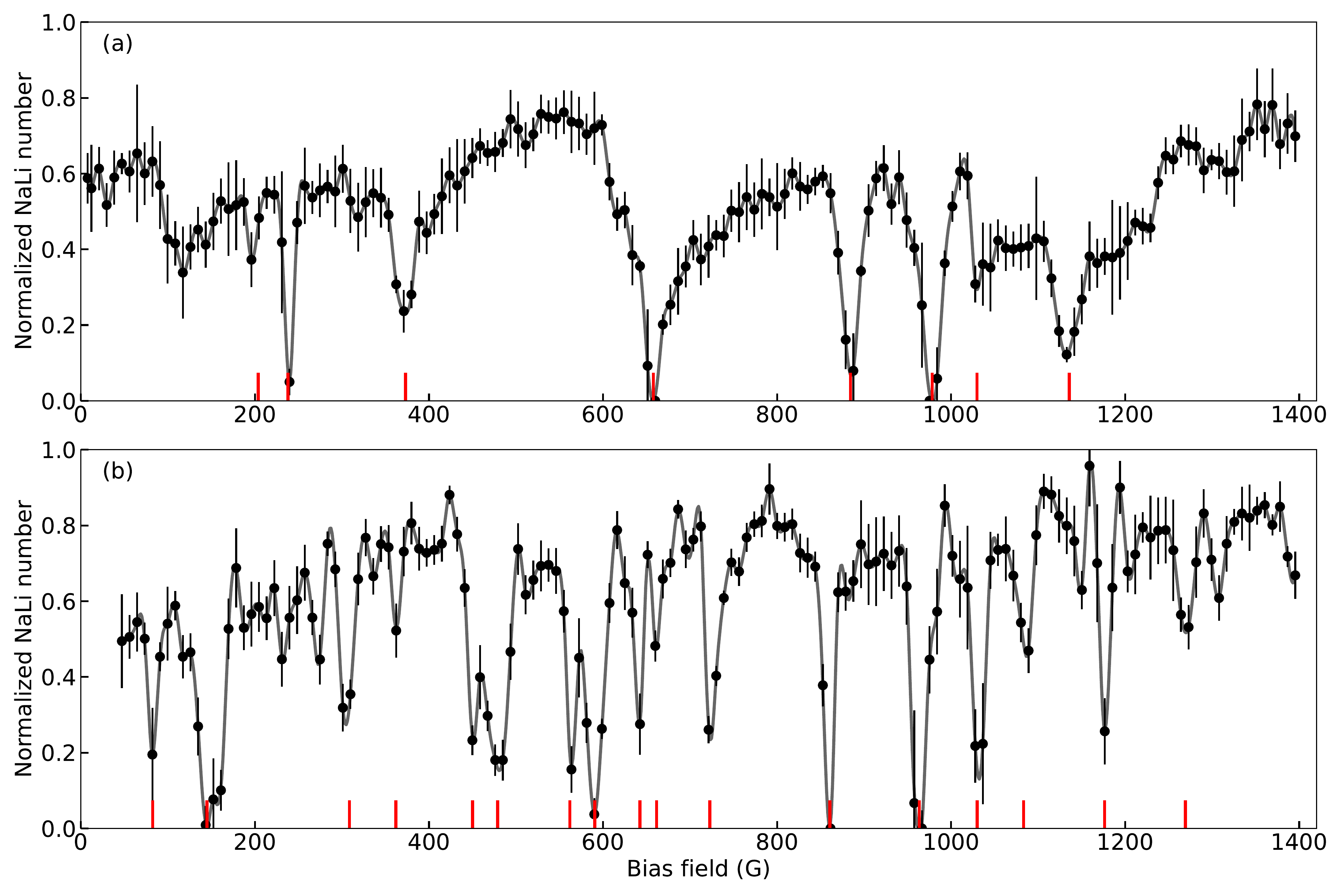}

	\caption{\textbf{Collisional loss spectrum of NaLi molecules with Na atoms as a function of magnetic field}. Spectra were recorded for both the upper (a) and lower (b) stretched hyperfine states.   Shown is the normalized number of NaLi molecules left after sweeping down the bias field by 13 G for 200 ms. The number of Na atoms in each pancake was around 115, and the initial number of NaLi molecules was about 35. A total of 25 resonances, indicated by red lines,  were observed: 8 in the upper spin-polarized mixture and 17 in the lower spin-polarized mixture.   Each data point represents three to eight measurements. Error bars correspond to one standard deviation of the mean. Grey lines are a guide to the eye, obtained by interpolating data with a piecewise cubic polynomial.}
	\label{fig:loss spectrum}
 \end{minipage}
\end{figure*}

\begin{table}[]
\setlength{\tabcolsep}{10pt} 
\renewcommand{\arraystretch}{1.5} 
\begin{tabular}{cccc}
\hline
\multicolumn{4}{c}{\textbf{Collision channel : $\mathbf{\ket{7/2,7/2}_{\rm{NaLi}} + \ket{2,2}_{\rm{Na}}}$}}          \\
 & $B_0 \; (\rm{G})$  & $\Delta B \; (\rm{G})$     & $\beta \; (\rm{cm}^3\rm{s}^{-1}$) \\ \hline
1 &  203.7(2)  & 5.3(7)    & $1.5(6)\times 10^{-11}$         \\
2 &  237.9(1)  & 6.3(4)    & $7.1(1) \times 10^{-11}$         \\
3 &  372.0(7)  & 20(3) & $2.3(5) \times 10^{-11}$         \\
4 &  657.8(3)  & 5.2(8)    & $1.0(1) \times 10^{-10}$         \\
5 &  884.3(3)  & 10(2)  & $6(1) \times 10^{-11}$         \\
6 &  978.2(2)  & 4.9(3)    & $7.4(5) \times 10^{-10}$         \\
7 &  1029.7(3) & 10(2)  & $1.5(3) \times 10^{-11}$         \\
8 &  1135.2(8) & 15(3) & $3.0(5) \times 10^{-11}$        \\
\hline
\hline
\\
\hline
\multicolumn{4}{c}{\textbf{Collision channel : $\mathbf{\ket{7/2,-7/2}_{\rm{NaLi}} + \ket{2,-2}_{\rm{Na}}}$}}          \\
 &  $B_0 \; (\rm{G})$  & $\Delta B \; (\rm{G})$     & $\beta \; (\rm{cm}^3\rm{s}^{-1}$) \\ \hline
1 &  82.5(1)    & 1.3(2)    & $1.0(3)\times 10^{-10}$ \\
2 &  145(1) & 18(4)   & $5.3(9)\times 10^{-11}$ \\
3 &  309(1)   & 13(4)   & $1.9(4)\times 10^{-11}$ \\
4 &  361.82(5)  & 1.0(2)    & $4.5(8)\times 10^{-11}$ \\
5 &  449.8(2)   & 3.1(5)    & $6(2)\times 10^{-11}$ \\
6 &  478(3)   & 35(13) & $3.4(6)\times 10^{-11}$ \\
7 &  561.8(2)   & 2.4(6)    & $8(2)\times 10^{-11}$ \\
8 &  590.7(2)   & 9(1)  & $8(2)\times 10^{-11}$ \\
9 &  642.3(5)   & 3(2)  & $2.9(4)\times 10^{-11}$ \\
10 &  661.5(2)   & 3.9(9)    & $2.4(5)\times 10^{-11}$ \\
11 &  722.5(1)   & 2.7(3)    & $8(4)\times 10^{-11}$ \\
12 &  860.8(2)   & 2.1(6)    & $2.4(6)\times 10^{-10}$ \\
13 &  963.3(2)   & 6(1)  & $2(1)\times 10^{-10}$ \\
14 &  1030.1(2)  & 3.3(8)    & $9(3)\times 10^{-11}$ \\
15 &  1083.3(3)  & 4(1)  & $5(1)\times 10^{-11}$ \\
16 &  1176.3(3)  & 3.2(9)    & $8(2)\times 10^{-11}$ \\
17 &  1269.2(1)  & 2.6(3)    & $3.3(5)\times 10^{-11}$ \\
\hline
\hline
\end{tabular}
\caption{\textbf{Observed Feshbach resonances.}  Resonance positions and widths are obtained by Lorentzian fits to loss features, except for resonances 4 to 6 for the incoming collision channel $\ket{7/2,7/2}_{\rm{NaLi}}+\ket{2,2}_{\rm{Na}}$. For these resonances, the position and width are determined by Lorentzian fits to the field-dependent loss rates \cite{son2022control}.}
\label{table:list of resonances}
\end{table}

We prepare a mixture of triplet ground-state $^{23}\text{Na}^{6}\text{Li}$$(a^{3}\Sigma ^{+})$ molecules and $^{23}\text{Na}$ atoms in the spin-polarized quartet potential. There are two possible states: the ``upper stretched state'', where all nuclear and electron spins are aligned to the bias magnetic field direction ($\ket{F, M_{F}}_{\rm{NaLi}} + \ket{F, M_{F}}_{\rm{Na}} = \ket{7/2, 7/2}_{\rm{NaLi}}+ \ket{2, 2}_{\rm{Na}}$) and the ``lower stretched state'', where all nuclear and electron spins are anti-aligned to the field direction ($\ket{7/2, -7/2}_{\rm{NaLi}}+ \ket{2, -2}_{\rm{NaLi}}$). Here, $F$ is the quantum number for the total spin (electron and nuclear) and $M_F$ is the $B$-field projection of $F$. The molecule and atom mixture in the upper stretched state with typical numbers of ${\sim} 3{\times}10^4$ and ${\sim}3{\times} 10^{5}$, respectively, is produced in a $1596$-nm one-dimensional optical lattice following the method described in Ref. \cite{NaLiSympCool, son2022control}.
The lower stretched state is produced by coherent transfer from the sample in the upper stretched state using a magnetic field sweep in the presence of radio frequency waves  \cite{park2023feshbach}. For this process, the bias field is dropped from $745$ G, where the upper stretched state is prepared, to a low field of around $8$ G in $15$ ms. After state preparation, the bias field is ramped to a target value in $15\; \text{ms}$.  Collisional lifetimes of the atom-molecule mixtures are determined by holding the sample for a variable time at the target magnetic field.

The loss of NaLi molecules with Na atoms is measured as a function of the bias field for both spin-polarized states. First, we performed a coarse search by sweeping the bias field over a range of 12.6 G during 200 ms and recording the number of remaining molecules, normalized to the number without the field sweep. This procedure was repeated over a range from near zero to 1420 G in steps of 8.76 G, with the results shown in Fig. \ref{fig:loss spectrum}.   We identified 8 resonances in the upper stretched state and 17 resonances in the lower stretched state indicated with red vertical lines in Fig. \ref{fig:loss spectrum}.

Near the resonances found in the coarse scan, finer scans were performed with sweep range and step size of around 1 G.  
\begin{figure}
    \centering
    \includegraphics[width = 83mm, keepaspectratio]{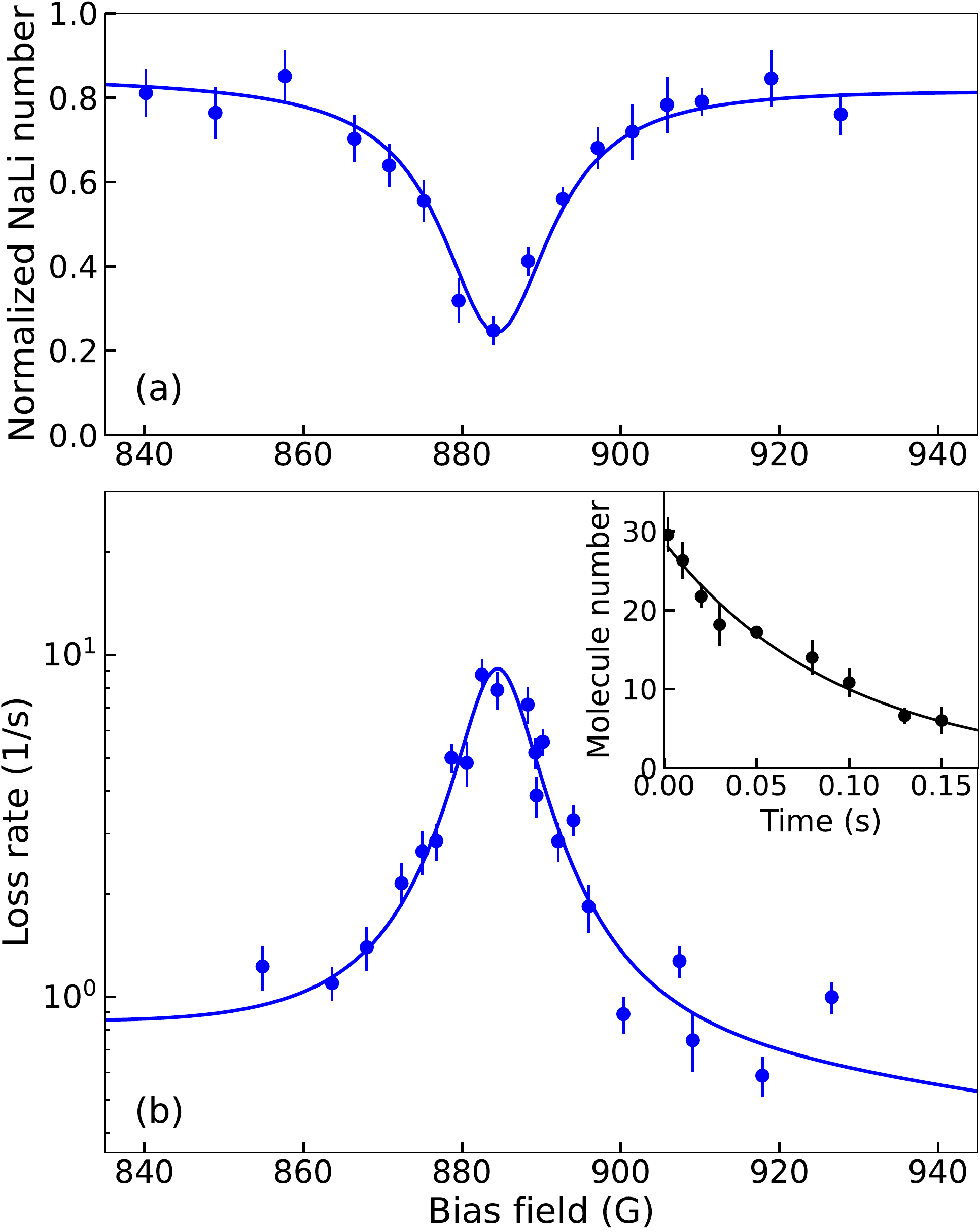}

	\caption{\textbf{Feshbach resonance near 884 G.}  This resonance was observed for NaLi + Na in the upper stretched state.  (a) Normalized NaLi number as a function of bias field after a hold time of 100 ms. (b) Loss rate of NaLi molecules obtained from decay curves as shown in the inset at 884.3 G. Data values and error bars in (a) and inset of (b) represent the average and one standard deviation of the mean, estimated from statistical errors of three to six measurements respectively. Error bars in the main plot of (b) represent one standard deviation of a fitted decay parameter. The blue lines are the fits to a Lorentzian function plus a background with a linear slope. 
	}
	\label{fig:resonance 5}
\end{figure}
Using the data from the fine scans, each loss feature was fit to a Lorentzian with a slope accounting for background loss and nearby resonances [an example is  shown in Fig. \ref{fig:resonance 5}(a)]. For all resonances, the thus obtained peak positions and widths are listed in Table \ref{table:list of resonances}.
Strong losses lead to a broadened lineshape. Therefore, for three resonances (labelled 4, 5 and 6 in the upper stretched state), we also determined the atom loss as a function of hold time and determined loss rates as a function of the bias field.  These results were used for a more accurate determination of the resonance position and width, as shown in Fig. \ref{fig:resonance 5}(b).
We confirmed that the width obtained from observed loss features can be broader than the width from the loss rate measurement. Fig. \ref{fig:resonance 5} illustrates this for resonance 5 in the upper stretched state which is near 884 G. The width obtained from a Lorentzian fit to the loss feature is 17(1) G, whereas the loss rate measurements give a width of 10(2) G. 
 
For all resonances, we determined peak loss rate constants $\beta$ by recording decay curves as a function of the hold time near the center of the loss features. They are summarized in the last column of Table \ref{table:list of resonances}. The determination of loss rate constants requires knowledge of the densities of sodium atoms overlapped with molecules. Instead of absolute calibration, we follow Ref. \cite{son2022control} and compare measured decay rates with the decay rate of the mixture in a non-stretched spin state which occurs at the universal rate~\cite{hermsmeier2021quantum}. This approach was validated in Ref. \cite{son2022control}.


Away from resonances, the observed background loss rates are more than an order of magnitude smaller than the universal loss rate constant which for Na $+$ NaLi $s$-wave collisions is $1.7 \times 10^{-10} \rm{cm}^{3} \rm{s}^{-1}$. More specifically, at low field near 0.5 G, the upper and lower stretched Na $+$ NaLi loss rate constants are $7.5(2.2) \times 10^{-12} \rm{cm}^{3} \rm{s}^{-1}$ and $6.7(2.0) \times 10^{-12} \rm{cm}^{3} \rm{s}^{-1}$, respectively, and at high field (near 540 G), they are $4.5(1.4) \times 10^{-12} \rm{cm}^{3} \rm{s}^{-1}$ and $3.5(1.0) \times 10^{-12} \rm{cm}^{3} \rm{s}^{-1}$, respectively.  Within the accuracy of measurement, the two stretched states have the same loss rate constants,  with lower background loss rates at higher magnetic fields.

\section{Analysis}
In this section, we summarize the experimental findings and draw some conclusions.  First, we have observed 
similar rates of background loss and Feshbach enhanced losses for both stretched states.   This implies that Zeeman energies do not play a major role and that dipolar relaxation is not the dominant decay mechanism.   Instead, the decay is probably dominated by shorter-range chemical reactions or inelastic loss, which are expected to be similar for both spin-stretched initial states.  Regarding the number of resonances, we have observed more resonances for the lower stretched state. In the range up to 1400 G, we observed 8 resonances in the upper stretched state and 17 in the lower stretched state. Intuitively this observation can be rationalized that for the lower stretched state,
the non-initial Zeeman states are closed channels that can support additional resonance states.
Surprisingly, our quantum chemistry calculations below show that the difference is instead caused by fast decay of some upper stretched state resonances that become too broad to be observed.

Next, we discuss the widths and spacings between resonances.  The linewidths of the resonances range from about 1 G to about 30 G. If the linewidths are interpreted as due to the finite lifetime of the resonant state, this implies atom-molecule complexes with lifetimes in the range of 10 to 350 ns.  The average spacing between resonances is around 100 G.  One possible interpretation of the spacing is rotational structure of the intermediate complex. The spacing between resonances of the order of 100 G corresponds to single spin-flip energy differences of 280 MHz.  A rotational constant of this value corresponds to the  moment of inertia of a complex with a size (i.e., atom/molecule separation) of around 30 $a_0$.  
Note that the rotationally excited states of the NaLi triplet molecules with $N =1$ are at 9.25 GHz and $N=2$ are at 27.75 GHz \cite{NaLiGround} corresponding to a double spinflip energy at 1650 G and 5000 G. In the accompanying paper, we show that this may lead to strong resonant features \cite{companion}, but these occur far outside the magnetic field range studied here.

Finally, we can perform a statistical analysis of the distribution of nearest-neighbor resonance spacings.  The observed distribution follows the Wigner-Dyson distribution \cite{wigner1993class, dyson1962statistical}, which is a feature of a chaotic system (see appendix). However, statistical conclusions from only 17 resonances are only tentative.

\section{Coupled-channels calculations}
With the coupled-channels calculations we will try to answer the following questions:
(1) What is the mechanism of coupling between the initial scattering channel and the loss channels for the background loss?  (2) What are the dominant interactions that are responsible for the resonances?  (3) Why are there more observed resonances for the lower stretched state?
(4) What quantum numbers describe the collisional complex?
(5) What is the size of the collisional complex?  

\begin{figure}
    \centering
    \includegraphics[width=\columnwidth]{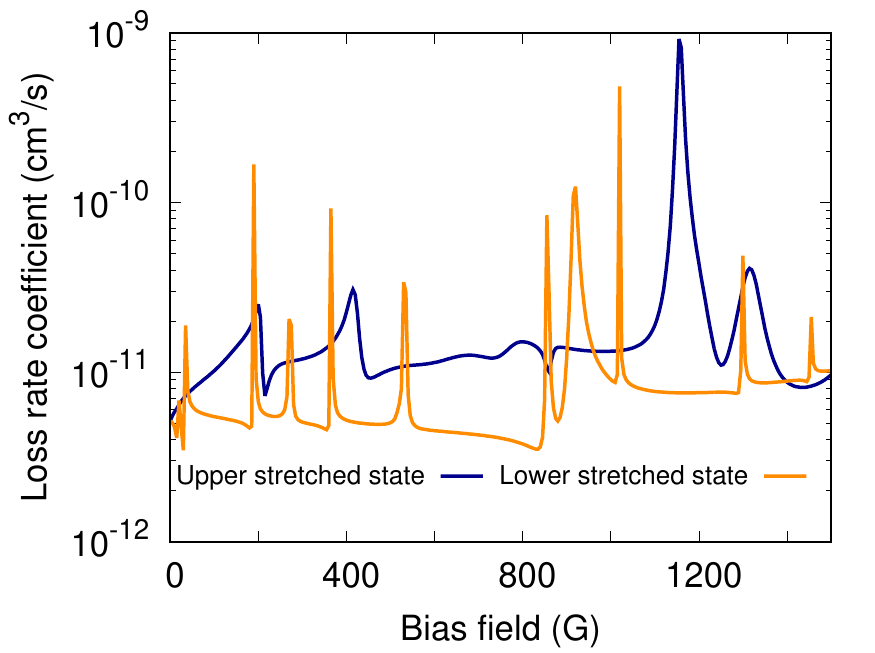}
    \caption{
\textbf{Calculated spectrum of Feshbach resonances}. Plotted are the calculated resonances for the upper and lower stretched states for $N_\mathrm{max}=30$ and $\lambda=-0.02$. The comparison with Fig. \ref{fig:loss spectrum} shows that most experimental features (except for resonance positions) are reproduced by the calculations.
	}
	\label{fig:thmain}
\end{figure}

The main result is shown in Fig.~\ref{fig:thmain}.
A calculated spectrum reproduces most of the observed features (magnitude of loss rate, widths and number of resonances, more resonances for the lower stretched states), but cannot predict the observed resonance positions. This illustrates the power (and limitations) of state of the art quantum chemistry calculations.

These calculations are fully quantum mechanical coupled-channel calculations including the electronic interactions, the Zeeman interaction with the magnetic field, the magnetic dipole-dipole interaction, and the spin-rotation and spin-spin couplings.
The electronic interaction is partitioned into two-body interactions that are taken from experiment \cite{knoop2011feshbach, steinke2012x, rvachov2018two},
and a non-additive three-body interaction that is calculated with state-of-the-art coupled-cluster methods including single, double, and triple excitations with large Gaussian basis sets extrapolated to the complete basis set limit that are described in the accompanying paper \cite{companion}. The interaction-induced variation of spin-rotation and spin-rotation couplings is neglected. We assume that the molecular bond length is fixed at the equilibrium position of the triplet potential, which naively cannot explain the chemical reactions that form singlet NaLi or Na$_2$ molecules at the low-spin potential. In our coupled-channel calculations, we model these by imposing an absorbing boundary condition at $R = 4.5~a_0$, which can be reached on the low-spin potential, but not on the high-spin potential, which is highly repulsive at these short distances.

For simplicity, we start by ignoring hyperfine and vibrational degrees of freedom. The scattering wavefunction is expanded in the basis of fully coupled channel functions of the form 
\begin{multline}
    \ket{(NL)J(\sm,\sa)S;\mathcal{J}\mathcal{M}} =\sum_{M_{J},M_{S}}{\braket{JM_{J}SM_{S}|\mathcal{J\mathcal{M}}}} \\
    \times \ket{(NL)JM_{J}}\ket{(\sm\ \sa)SM_{S}}
\label{eq:basis 1}
\end{multline}
where $\braket{JM_{J}SM_{S}|\mathcal{J\mathcal{M}}}$ is a Clebsh-Gordan coefficient. The quantum number $N$ represents the rotational angular momentum of the NaLi molecule, and $L$ the angular momentum associated with the relative motion of the atom and molecule.
$N$ and $L$ are Clebsch-Gordan coupled to a total mechanical angular momentum $J$ with $z$-component $M_J$.
Similarly, $\sm=0$ or $1$ denotes the NaLi molecular electronic spin and $\sa= \half$ is the atomic electronic spin, and $S$ the total electronic spin with $B$-field projection $M_S$. In the coupled basis, $J$ and $S$ are subsequently coupled to a total angular momentum $\mathcal{J}$ and a magnetic field projection $\mathcal{M}= M_J+M_S$. $\mathcal{M}$ is strictly conserved, whereas, for a large enough magnetic field, $M_S$ becomes a good quantum number and therefore $M_J = \mathcal{M}-M_S$ is also a good quantum number.
Due to the large singlet-triplet splitting in the NaLi molecule, $\sm = 0$ or 1 is also a good quantum number.
For a separated atom and molecule, $m_{\sm}$ and $m_{\sa}$ would separately become good quantum numbers, but at chemically relevant distances the exchange splitting between the doublet and quartet interaction potentials is dominant, so that $S = 1/2$ and $3/2$ are good quantum numbers.
Hence, we can effectively consider each $\ket{S\ M_S}$ state separately, with only perturbatively weak couplings between them.
For each of these spin channels, there are strong and anisotropic interactions that couple different $N$ and $L$ channels but conserve $J$ and $M_J$.
Since we ignore nuclear spin, the initial channel corresponds to $s$-wave collisions in the upper spin-stretched ground state, 
$\ket{(NL)JM_{J}}\ket{SM_{S}} = \ket{(00)00}\ket{3/2\; 3/2}$,
or the lower spin-stretched state $\ket{(00)00}\ket{3/2,\; -3/2}$.

\begin{figure}
    \centering
    \includegraphics[width = 83mm, keepaspectratio]{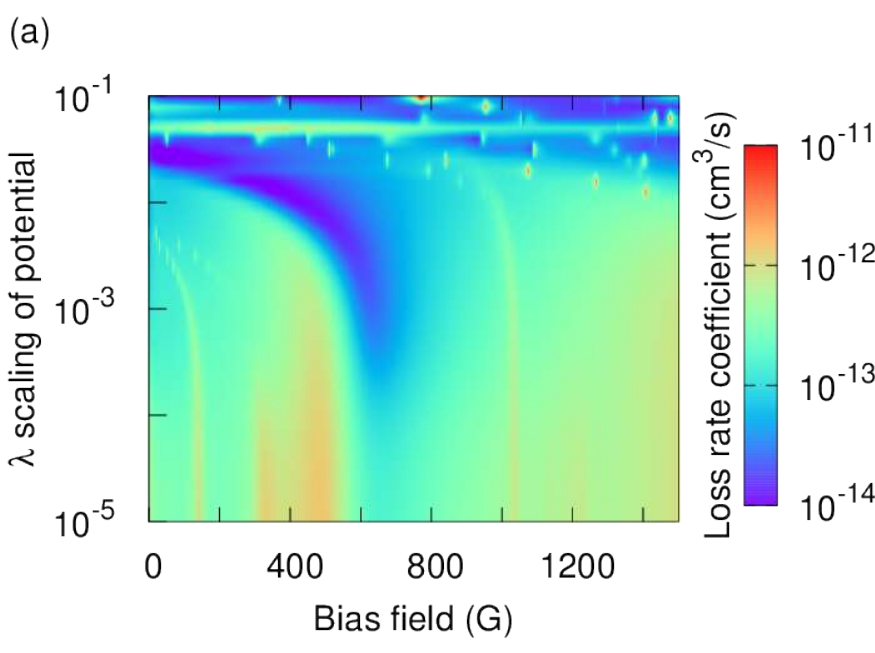}
    \includegraphics[width = 83mm, keepaspectratio]{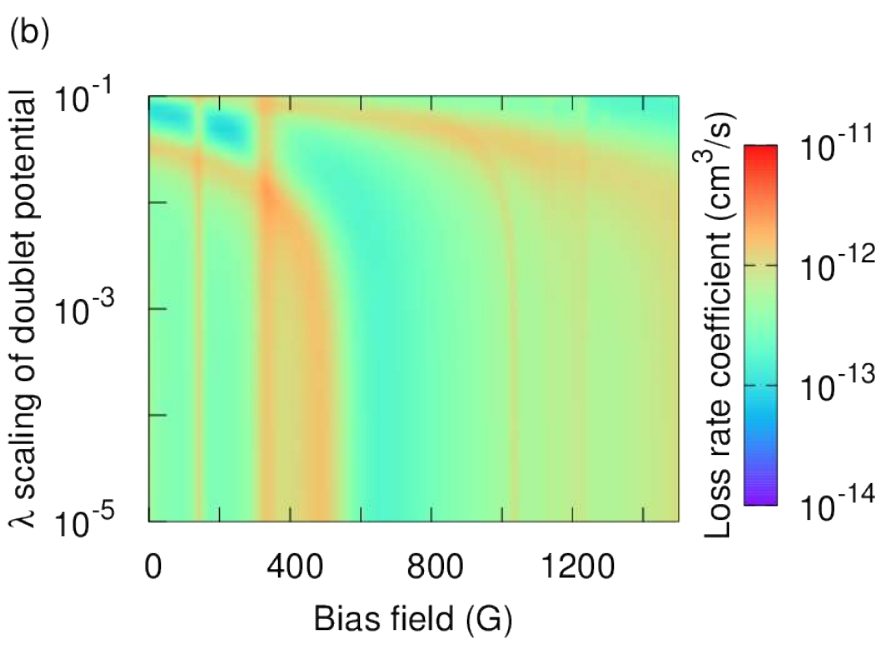}
    \caption{\textbf{ Calculated loss rates as a function of magnetic field.}   Calculations are done with scaling of the spin-independent three-body interaction by a factor $1+\lambda$ for $N_{\rm{max}} = 6$. (a) Scaling of the three-body interaction for both the doublet and quartet state.  The figure shows that scaling the three body interactions within their uncertainty of at least several percent dramatically changes the loss rates:  The positions of several Feshbach resonances change, and several $B$-independent resonances appear when the $\lambda$-scaling creates a resonance in the initial spin-stretched potential, i.e., a bound state near zero energy. The conclusion is that the prediction of resonance positions and background loss requires knowledge of the interaction potentials to an accuracy better than can be achieved by \emph{ab initio} calculations. (b) Scaling of the three-body interaction only for the doublet potential. In this case, the resonances follow vertical lines and  are independent of the scaling up to $\lambda= 0.1$, implying that the resonances originate in the quartet potential.
	}
	\label{fig:lambda dependence}
\end{figure}

First, we consider the sensitivity of the scattering rates to the interaction potential shown in Figure~\ref{fig:lambda dependence}(a).
Here, we scale by a factor $1+\lambda$ the non-additive three-body part of the interaction potential, that is, the part that is computed \emph{ab initio} and is uncertain up to an conservatively estimated 5~\%.
By modifying the potential by only 0.1~\% we find that the resonances start to shift so that realistically their positions are completely undetermined,
and when the scaling reaches several percent we tune across magnetic field-independent resonances, indicating that the background scattering length is undetermined.
Next, shown in Figure~\ref{fig:lambda dependence}(b), we again scale the three-body interaction but now only for the low-spin doublet potential,
leaving the high-spin quartet potential unchanged.
In this case, we find that several of the resonances are now completely independent of the scaling of the low-spin potential up to $\lambda= 0.1$.  This implies that no stable resonance states are supported by the chemically reactive low-spin potential and that all the predicted (and observed) resonances originate in the quartet potential.

The analysis above indicates that the \emph{ab initio} prediction of resonance positions is beyond the capability of state-of-the-art calculations.
Although the coupled-channels calculations cannot predict the positions of the resonances,
we can still use these calculations as a ``numerical experiment'' to investigate the nature of the resonance states, the coupling mechanisms, and the observed differences between the two spin-stretched states.

\begin{figure}
    \centering
    \includegraphics[width=\columnwidth]{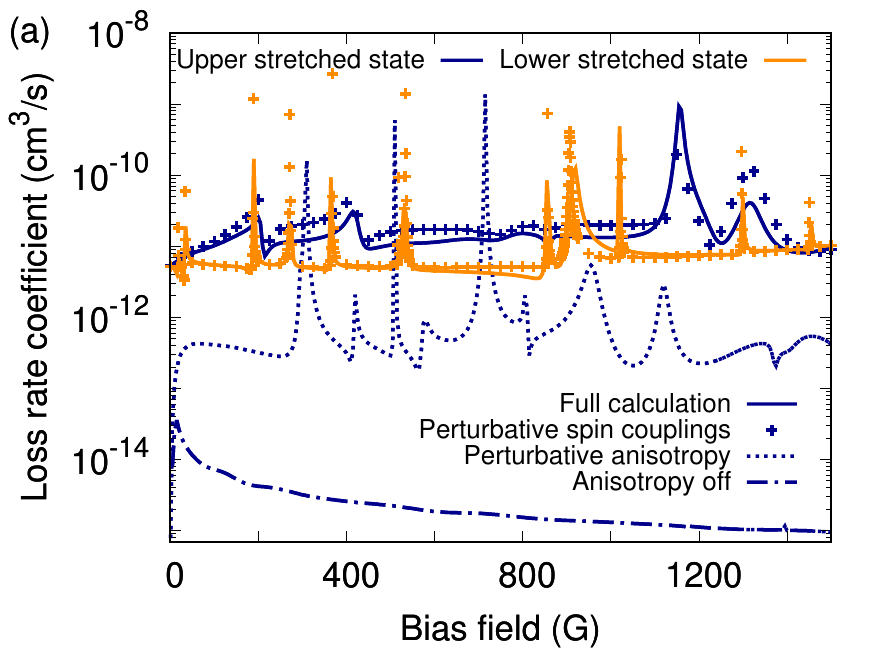}
    \includegraphics[width=\columnwidth]{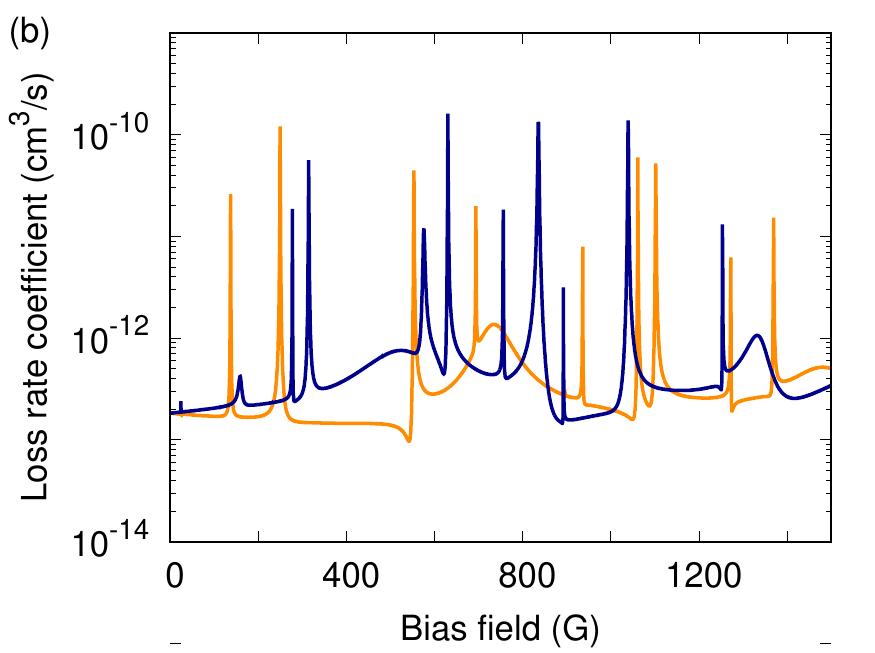}
    \caption{
\textbf{Calculated spectrum of Feshbach resonances.}  Figure (a) shows results for the upper and lower spin-stretched states in orange and blue solid lines (same data as in Figure \ref{fig:thmain}).
The lower spin-stretched state typically shows around 10 resonances below 1500~G,
whereas the upper spin-stretched state shows around half as many resonances.
This qualitative difference between the two states is also observed experimentally.
We investigate whether the anisotropy of the electronic interaction and the spin-spin and spin-rotation coupling act perturbatively, by scaling down these couplings by a factor of two and scaling the resulting cross section up by a factor four.
Agreement with the full calculation indicates the spin-spin and spin-rotation coupling act perturbatively,
whereas the interaction anisotropy does not.
The coupling mechanism however does involve the anisotropy as turning this off entirely produces a much smaller cross section dominated by the magnetic dipole-dipole interaction (dash-dotted line).
When both anisotropy and dipole-dipole are turned off, the calculated cross section is zero.
(b) Removal of Zeeman states.  The qualitative differences in the number of resonances for the upper and lower spin-stretched states disappear when we \emph{exclude} from the calculation channels corresponding to Zeeman states in the rotational ground state, except for the initial one. 
	}
	\label{fig:rep}
\end{figure}

We perform coupled-channels scattering calculations with the interactions scaled by $1 +\lambda$ and analyze the \emph{typical} behavior observed for different $\lambda$ between $-0.1$ and $+0.1$.
Representative magnetic field scans can be seen in Fig.~\ref{fig:rep}(a).
We observe approximately 10 resonances for the lower spin-stretched state and only around 5 resonances in the upper spin-stretched state, respectively.
This is in qualitative agreement with the experiment which observes 17 and 8 resonances, respectively.
In the companion paper we show that based on the density of states we would expect to see approximately 10 resonances for either spin-stretched state \cite{companion}.
We furthermore show that including molecular vibrations would increase the density of state by approximately 50~\%,
which can partially explain the lower number of observed resonances compared to experiment.
We show that including hyperfine interactions increases the density of resonances somewhat.
With these effects in mind, one could  claim almost quantitative agreement with experiment regarding the density of resonances.

One may expect that the lower spin-stretched state supports more magnetically tunable resonances because the non-initial Zeeman states correspond to closed channels -- and hence can support Feshbach resonances -- even in the rotational ground state,
whereas closed channels for the upper spin-stretched state occur only for excited rotational states.
We investigate this in our coupled-channels calculations by artificially excluding the non-initial Zeeman states in the rotational ground state, see Fig.~\ref{fig:rep}(b).
Somewhat surprisingly we find that \emph{excluding} channels from the calculation does not reduce the number of resonances for the lower spin-stretched state,
but rather increases the observed number of resonances for the upper spin-stretched state, where the excluded channels correspond to asymptotically open channels.
In the presence of these open channels some of the resonances decay rapidly by spin-spin coupling to lower Zeeman levels (both in the doublet and quartet potentials) such that they are not resolved, leading to a lower number of observable resonances. This explains the observed qualitative difference between the upper and lower spin-stretched states.

Finally, we investigate numerically the coupling mechanism that gives rise to the observed resonances.
As argued above, each resonance state can essentially be assigned a molecular and a total electron spin quantum number $\sm=1$ and $S={}^3/_2$,
as only the non-reactive quartet spin state supports stable resonance states.
The resonance is magnetically tunable only if the Zeeman state $M_S$ changes.
To couple states with $\Delta M_S \neq 0$ a spin-dependent coupling must be involved through spin-rotation and spin-spin coupling.

From the tensor rank of these couplings we can determine they couple states with the selection rules,
$J = 0 \rightarrow 1$ and $|\Delta M_S| \le 1$ (for spin-rotation coupling), and $J = 0 \rightarrow 2$ and $|\Delta M_S| \le 2$ (for spin-spin coupling), respectively.  The spin-spin couplings and spin-rotation couplings arise from terms $(\hat{s}_\mathrm{mol} \cdot \hat{r}) (\hat{s}_\mathrm{mol} \cdot \hat{r})$ and $\hat{s}_\mathrm{mol} \cdot \hat{N}$ for the NaLi molecule, where $\hat{r}$ points along the molecular axis and $\hat{s}_\mathrm{mol}$ is the molecular electronic spin.  Since they depend on the orientation of the molecular axis, they cause exchange of spin angular momentum with the molecular rotation $N$.

Since $J$ is approximately a good quantum number, both mechanisms give rise to distinct and independent resonances.
Since the density of states of the collision complex scales with $2J+1$ and the differential magnetic moment is higher for larger $|\Delta M_S|$ transitions,
we conclude that most -- approximately 5/6 -- of the resonances are due to spin-spin coupling,
and the remaining 1/6 is due to spin-rotation coupling.
Spin-spin coupling does not occur for $^2\Sigma$ molecules, and hence is a somewhat unique coupling mechanism for NaLi in the triplet ground state ($^3\Sigma$).

To confirm the role of the spin-dependent interactions in the coupling mechanism we have performed coupled-channels calculations where we reduced these couplings by a factor two.
The resulting cross section multiplied by four is shown as the crosses in Fig.~\ref{fig:rep}(a).
The agreement with the full calculation indicates scaling with the square of the coupling strength that is expected for perturbatively weak spin-dependent couplings. 
Note that on resonance, the dependence on the coupling strength is not an overall scaling,
as the coupling strength also determines the resonance widths.
For smaller coupling, the peaks are narrower and higher.
Those results confirm that we can fully assign the resonances approximately good quantum numbers $\sm=1$, $S={}^3/_2$, each resonance also has definite $M_S$ constrained by $\Delta M_S \le 1$ and $\le 2$ selection rules, and $J=1$ or $2$, for spin-rotation and spin-spin coupling, respectively,
whereas the $N$ and $L$ quantum numbers are strongly mixed due to the anisotropic interaction at short range.
Figure~\ref{fig:rep}(a) also implies that the loss mechanisms are the same for resonant losses and background losses.  For the value of $\lambda$ chosen in Fig.~\ref{fig:rep}(a), the background loss is higher for the upper spin-stretched state, but for most values of $\lambda$, they are similar, as observed in the experiment~\cite{companion}.

The coupling mechanism involving the spin-rotation and spin-spin coupling also requires an anisotropic interaction potential.
The physical picture is that the anisotropic interaction with the atom can reorient the molecule,
and because the spin is coupled to the molecular axis by spin-rotation and spin-spin coupling, this can lead to Zeeman transitions.
To confirm this picture we have performed calculations that exclude interaction anisotropy, shown as the dash-dotted line in Fig.~\ref{fig:rep}(a).
The resulting cross section is much smaller and results from long-range Zeeman relaxation by the magnetic dipole-dipole coupling.
If both the magnetic dipole-dipole interaction and the interaction anisotropy are switched off, the cross section in our model vanishes.
The role of the interaction anisotropy is non-perturbative, however, as can be seen from comparison between the solid and dotted line in Fig.~~\ref{fig:rep}(a),
which compares the cross section from the full calculation to four times the cross section obtained with the interaction anisotropy halved.
The crucial and non-perturbative role of the anisotropic electronic interaction implies that the spectrum of resonances cannot be described by a simplified model that accounts only for the isotropic long-range $R^{-6}$ interaction,
contrary to previous observations of ultracold atom-molecule resonances \cite{wang2021magnetic}.
In summary: The observed losses arise from the interplay of spin-spin and spin-rotation coupling with the anisotropic interactions which connect the incoming non-reactive quartet states to doublet states that can decay to singlet NaLi or Na$_2$ molecules. This coupling can occur directly, or via first coupling  to non-stretched quartet states.

We make a direct comparison to the scattering calculations with the density of states computed quantum mechanically using the same channel basis as used in the scattering calculations in the accompanying paper \cite{companion}. The total number of states from the quasiclassical estimate is close to the number of resonances from the scattering calculations, but not in perfect agreement with it, because of the light masses and relatively weak interactions in the spin-stretched state. We find that most of the 3-body states are between 20 and 40~$a_0$ which is in agreement with the complex size estimated from the simple interpretation of the spacing between resonances as rotational energy splitting of the collision complex. This size is short compared to the range of the van der Waals potential, but much longer than the short-range interactions - the minimum of the potential wells is around 10~$a_0$.  The resonances depend both on short-range and long-range physics: They are supported by the long-range potential, but require the anisotropic short-range interactions as a coupling mechanism.

The calculated widths of the resonances vary between 1 and 30 G, in qualitative agreement with the experiment.  Those widths reflect the lifetime since there are no other broadening mechanism in our simulations. Experimental broadening such as finite resolution of magnetic field strength and magnetic field inhomogeneity is very small (estimated to be around 100 mG near 1400 G).

\section{Conclusion and Outlook}
We have provided a large-scale map of Feshbach resonances between ultracold atoms and molecules. We have chosen collisions between NaLi in the triplet ground state and Na for the two fully spin-stretched initial collision channels in the quartet potential.  Although the NaLi $+$ Na mixture is chemically reactive, there is no reactivity in the quartet potential, i.e. inelastic collisions require a spinflip to the doublet potential  \cite{son2022control, tomza2013chemical}.  The possibility of metastable collision complexes makes this system promising for the study of Feshbach resonances.  Indeed, we have observed 8 and 17 resonances within a $\sim$1400 G range in the upper and lower stretched states, respectively. 



We have compared the experimental results to full coupled-channel calculations for the NaLi$+$Na mixture. Even state-of-the-art quantum chemistry calculations cannot predict the position of resonances because of the uncertainty in the interaction potentials. However, they can provide a deep understanding of the relevant states and their couplings. 
Our simulations have shown that the resonance states are 3-body complexes in the quartet state with rotational excitation, either of NaLi (quantum number $N$), or rotation of Na around NaLi (quantum number $L$). These complexes have a typical size of 30-40~$a_0$. The input state and the complex state involve different $M_S$, which is necessary for a magnetic Feshbach resonance. Molecular eigenstates in different Zeeman levels can be coupled by a strong anisotropic electronic interaction since the different molecular Zeeman states have different rotational state decompositions mainly due to spin-spin coupling and also spin-rotation coupling. Note that the spin-rotation and spin-spin coupling depend on the orientation of the molecular axis and, therefore, provide tensorial couplings between the spin and the mechanical angular momentum. Approximately 5/6 of the resonances can be attributed to spin-spin coupling to states with total mechanical angular momentum $J=2$, and the remaining 1/6 is attributed to spin-rotation coupling and assigned $J=1$.

The loss of molecules, either the background loss or via Feshbach resonances, occurs through transition  to the chemically reactive doublet potential, either directly or through lower Zeeman states of the quartet potential (for the upper stretched state).  These transitions involve an interplay of the anistotropic electronic interaction with spin-spin or spin-rotation coupling in the NaLi molecules. Dipolar transitions are considerably weaker.

Our coupled-channels calculations explain why the number of observed resonances is smaller for the upper stretched state. The upper stretched state supports a similar number of resonances, but several of them are washed out by the rapid decay of the intermediate complexes to lower-lying Zeeman states via spin-spin coupling.

The agreement between experiment and calculations validates the assumptions made in the calculations.  At least for a semi-quantitative analysis, it is sufficient to neglect hyperfine interactions and vibrationally excited state of the molecules, and use electronic potentials at fixed nuclear separation for the molecule.

The physical mechanisms we have identified in the Na $+$ NaLi($a^3\Sigma^+$) mixture are unique to molecules in the triplet or higher-spin state.  
It would be interesting to study other bi-alkali triplet molecules and check if their collision dynamics is dominated by similar couplings. The only other case where atom-molecule Feshbach resonances have been studied involved NaK in the singlet ground state  \cite{yang2019observation, wang2021magnetic}. For this system, the observed Feshbach resonances were caused by long-range van der Waals interactions \cite{wang2021magnetic,frye2021complexes}.
These long-range states then follow regular patterns dictated by the quantum defect theory for a simple isotropic long-range $R^{-6}$ interaction \cite{Frye:22}.
The resonances described here, however, behave quite differently as the coupling mechanism involves perturbatively the spin-spin and spin-rotation coupling,
and the strong non-perturbative anisotropic interaction at short range.
Although the resulting resonances are supported by the long-range interaction,
their positions are sensitive to this anisotropic short-range interactions and hence cannot be described by the regular patterns predicted by quantum defect theory.

Our work here has focused on the non-reactive quartet potential to support metastable collisional complexes.  
However, recent studies have shown that collisional resonances and collisional complexes should also occur in highly reactive systems. A single $p$-wave Feshbach resonance has been observed in collisions between NaLi molecules in the triplet ground state \cite{park2023feshbach}.  This system has no barrier for reactions.  Losses much smaller than the universal loss rate have been observed in s-wave collisions between magnetically trapped NaLi molecules in different hyperfine states of the triplet state \cite{park2022magnetic}, a system which should be highly reactive.  Losses far below the universal rate imply a finite reflection probability at short range and should lead to Feshbach resonances when the magnetic field tuning creates strong interference between reflections at short and long range  \cite{son2022control, PJ_universal}. 

These studies emphasize that our understanding of reactivity and metastable collisional complexes is incomplete, and further experimental explorations are required.  Further studies should involve NaLi in various hyperfine states and with various collision partners, as well as considering other bi-alkalis. As suggested in Ref.~\cite{frye2021complexes}, for non-stretched states, the Fermi contact interaction between electron and nuclear spin provides another mechanism for Feshbach resonances.  Such broader studies are required to find out which mechanisms are universal, and which occur only in specific systems.  

\begin{acknowledgments}
The MIT work was supported by the NSF through the Center for Ultracold Atoms and Grant No. 1506369 and from the Air Force Office of Scientific Research (MURI, Grant No. FA9550-21-1-0069). J. J. P. and H. S. acknowledge additional support from the Samsung Scholarship.  M. G. and M. T. were supported by the National Science Centre Poland (Sonata Bis Grant no. 2020/38/E/ST2/00564) and the PL-Grid Infrastructure (computational grant no. PLG/2020/014342).

\end{acknowledgments}


    

\appendix	
\section{Statistical analysis}

\begin{figure}
    \centering
    \includegraphics[width = 83mm, keepaspectratio]{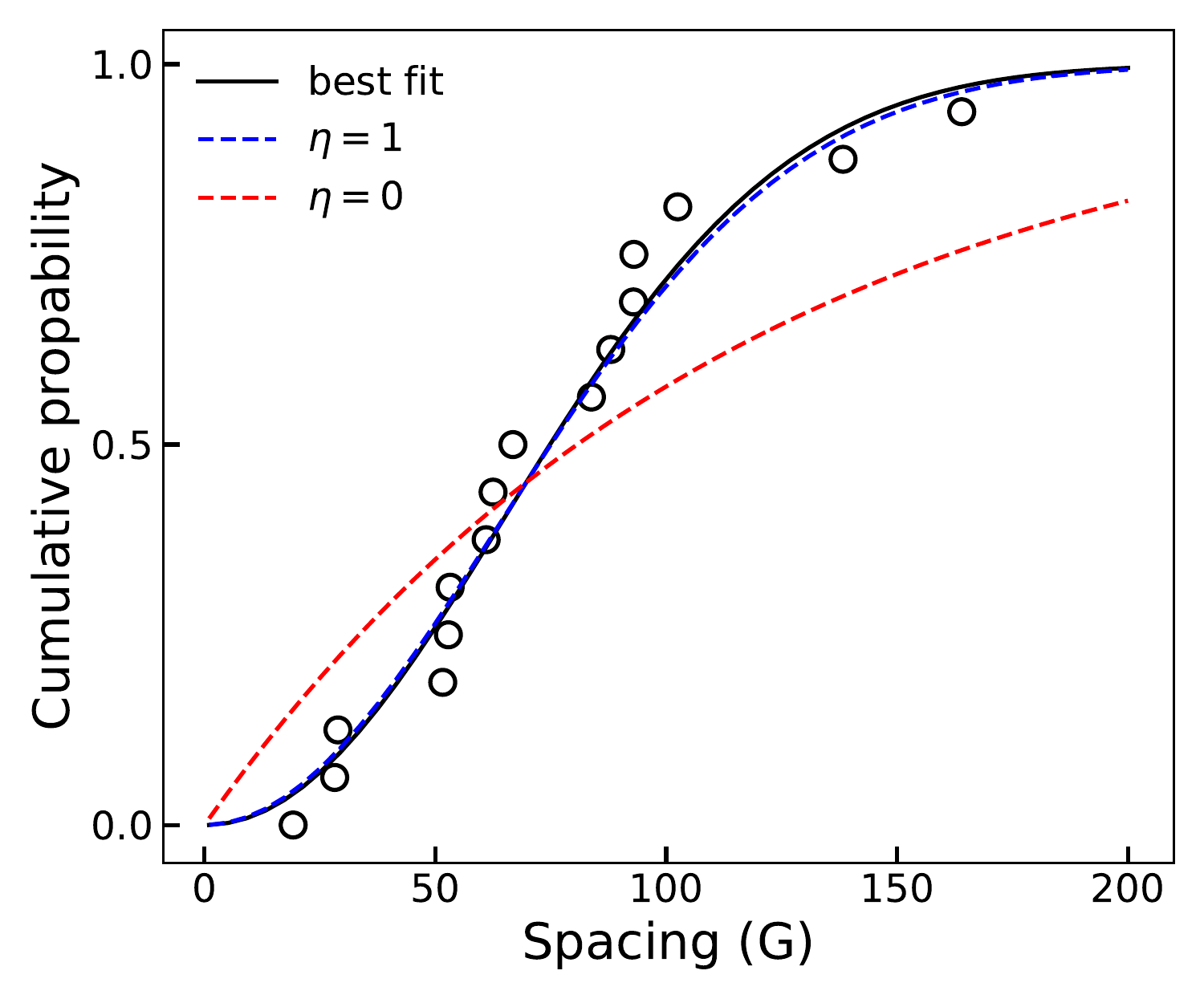}
	\caption{{\bf  Cumulative probability of resonance spacings.}  Shown are the experimental results for Na+NaLi collisions in the lower stretched state. Black line is the best fit to the cumulative probability function of the Brody distribution (Eq.\ref{eq:Brody}) and the blue (red) dashed line is the cumulative distribution with $\eta=1$ ($\eta=0$).
	}
	\label{fig:Brody cumulative distribution}
\end{figure}

A statistical analysis of the separations between resonances can provide insight into the nature of the resonant states.  We quantify the resonance statistics of 17 resonances from Na+NaLi collisions in the lower stretched state provided in Table \ref{table:list of resonances} using the Brody parameter $\eta$, which is a standard measure of chaos. For non-chaotic systems, in which resonances have no correlations, the distribution of nearest-neighbor spacings is given by the Poisson distribution, $P_{\text{P}}(s)=e^{-s}$. On the other hand, for chaotic systems, which emerge when the mean spacing between bound states is comparable to the coupling strength, repulsion between energy levels occurs. In this regime, the distribution of nearest-neighbor spacings is given by a Wigner-Dyson distribution, $P_{\text{WD}}(s)=\frac{\pi}{2}se^{-\frac{\pi}{4}s^{2}}$ \cite{wigner1993class, dyson1962statistical}. The two distributions are smoothly interpolated through the Brody parameter $\eta$ as $P_{\text{B}}(s)=As^{\eta}e^{-\frac{A}{\eta+1} \cdot s^{\eta+1}} $, known as the Brody distribution, where $A=(\eta+1)\cdot\Gamma\left(\frac{\eta+2}{\eta+1}\right)^{\eta+1}$ \cite{brody1973statistical, brody1981random}. Here, $\eta =1$ and $\eta=0$ lead to the Wigner-Dyson and Poisson distributions, respectively. The cumulative probability function of the Brody distribution is given as

\begin{equation}
    F_{B}(s)=(\eta+1)\cdot[1-e^{-\alpha s^{\eta+1}}].
\label{eq:Brody}
\end{equation}
Fig. \ref{fig:Brody cumulative distribution} shows the cumulative probability of resonance spacing of the 17 resonances from Na$+$NaLi collisions in the lower stretched state. The best fit of the data to Eq.\ref{eq:Brody} gives $\eta=1.1(1)$, which shows the statistical signature of quantum chaos.

\begin{figure}
    \centering
    \includegraphics[width = 83mm, keepaspectratio]{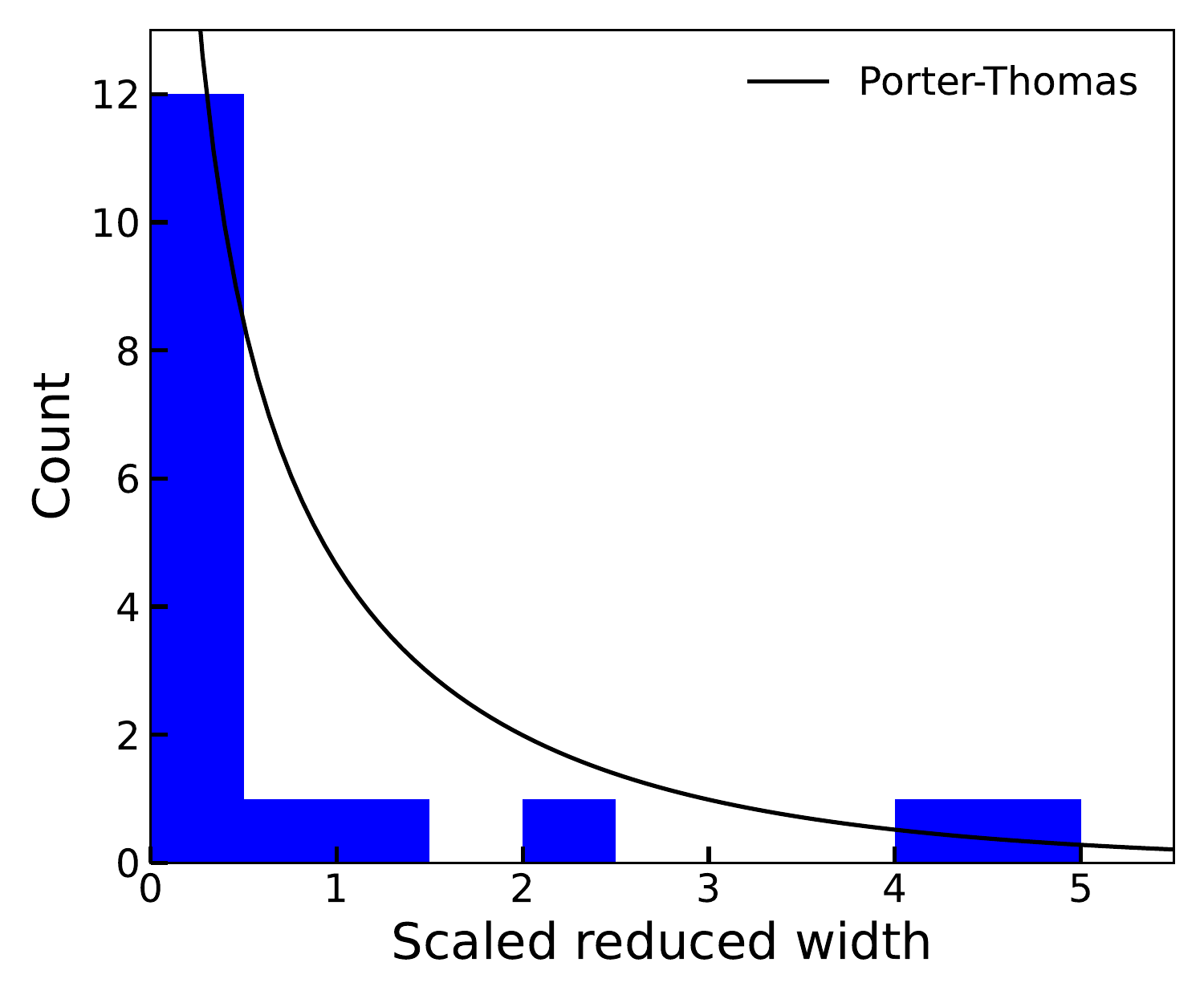}
	\caption{
	{\bf Distribution of scaled reduced widths for scattering resonances.} This plot is based on the data for the lower stretched state Na+NaLi mixture listed in Table \ref{table:list of resonances}. The black line shows the Porter-Thomas distribution of scaled reduced widths.  
	}
	\label{fig:Porter Thomas}
\end{figure}

However, statistical conclusions from only 17 resonances are tentative. On the basis of our quantum-scattering calculations, the Wigner-Dyson statistics is not expected because each resonance can be assigned by $M_S$, $J$, and $M_J$ quantum numbers (resonances with different values for these quantum numbers do not affect each other). In addition, it is difficult to conclude that the broad trend of the distribution of resonance widths follows the Porter-Thomas distribution of resonance widths, $P_{\text{PT}}(\bar{\gamma})=\bar{\gamma}^{-1/2}e^{-\bar{\gamma}/2}$, which is also a statistical character of quantum chaos. Here, the scaled reduced widths, $\bar{\gamma}=\gamma_n^0 / \langle \gamma_n^0 \rangle$ where $\gamma_n^0$ is the resonance width normalized by the resonance energy, and $\langle\gamma_n^0 \rangle$ is the average of $\gamma_n^0$ \cite{porter1956fluctuations}. Fig. \ref{fig:Porter Thomas} shows the histogram of the scaled reduced resonance width distribution and the Porter-Thomas distribution. Due to the small statistical sample, it is difficult to compare the broad trend of the distribution with the Porter-Thomas distribution. Nevertheless, Fig. \ref{fig:Brody cumulative distribution} represents a purely empirical analysis of experimental data, which should find a theoretical explanation (which is probably not quantum chaos).
\bibliography{main}

\end{document}